\documentclass[a4paper,aps,onecolumn,nofootinbib]{revtex4}
\RequirePackage[colorlinks,hyperindex]{hyperref}
\RequirePackage[english]{babel}
\RequirePackage[latin1]{inputenc}
\RequirePackage[T1]{fontenc}
\RequirePackage{mathrsfs}
\RequirePackage{amsmath}
\RequirePackage{amssymb}
\RequirePackage{amsbsy}
\RequirePackage{color}
\RequirePackage{bm}
\hypersetup{colorlinks=true,breaklinks=true,urlcolor=blue,linkcolor=red}
\begin{document}
\title{\bf{Dirac hydrodynamics in 19 forms}}
\author{Luca Fabbri\footnote{luca.fabbri@edu.unige.it}}
\affiliation{DIME, Universit\`{a} di Genova, Via all'Opera Pia 15, 16145 Genova, ITALY\\
INFN, Sezione di Genova, Via Dodecaneso 33, 16146 Genova, ITALY}
\date{\today}
\begin{abstract}
We consider the relativistic spinor field theory re-formulated in polar variables so to allow for the interpretation given in terms of fluid variables. After that the dynamics of spinor fields is converted as dynamics of a special type of spin fluid, we demonstrate that such conversion into dynamical spin fluid is not unique but it can be obtained through $19$ different rearrangements, by explicitly showing the $19$ minimal systems of hydrodynamic equations that are equivalent to the Dirac equations.
\end{abstract}
\maketitle
\section{Introduction}
In mathematics, writing complex functions as a product of a module times a unitary phase, that is writing a complex function in its polar decomposition, is a well known method extensively used to treat a variety of problems.

In physics, this method can be applied whenever complex functions are involved, and of all cases, the most important is certainly quantum mechanics. In its non-relativistic spinless case, the wave function is a complex scalar, so that its polar decomposition is straightforward. When the wave function written in polar form is inserted into the Schr\"{o}dinger equation, this last splits in a continuity equation for the velocity density plus a quantum mechanical extension of the Hamilton-Jacobi equation. This procedure was first applied by Madelung, and it is widely known today. As in their final form the field equations are expressed as derivatives of quantities like density and velocity, this procedure actually converts quantum mechanics into a special type of fluid mechanics, and since the continuity equation states that this fluid must be incompressible, such a formulation of quantum mechanics is called hydrodynamic formulation \cite{DB,TT}.

A more realistic description of quantum mechanics can be given when it is extended so to include spin. Still within the validity of non-relativistic regimes, the wave function is now a complex doublet that transforms in a specific way under rotations, but a polar form can be obtained just the same. After the inclusion of spin the Schr\"{o}dinger equation is known to be enlarged into the Pauli equations, but again when the spinorial wave function written in polar form is used, a hydrodynamic formulation becomes possible \cite{Bohm-Schiller,Takabayasi1954}. As it stands, one would now assume that in the relativistic case, in which the wave function is formed by two complex doublets transforming in a very specific way under boosts and rotations, an analogous polar form could also be obtained. And since in the relativistic case the Pauli equations are replaced by the Dirac equations, one should also expect the relativistic spinorial field in polar form to convert the Dirac equations into the corresponding hydrodynamic formulation \cite{B-T,Takabayasi1957}. This is exactly what will happen eventually although not with the immediate chronology that one might have guessed. And the reason is covariance.

In fact, in the passage from a wave function of scalar character to that of spinor character, whether non-relativistic or relativistic, there are additional transformation laws that have to be considered. As already mentioned, non-relativistic spinors are doublets that account for both helicity states, and while each component is a scalar for diffeomorphisms (that is for passive transformations of coordinates), the two components mix under rotations (that is for active changes of frame). Richer in structure, relativistic spinors are columns of four components that account for both chiral states as well as both helicity states, and while again each component is a scalar for diffeomorphisms, the four components mix under Lorentz transformations. It is now easier to see where the difficulty may be found. Indeed, considering the relativistic spinor, the passage to its polar form would be implemented when, for the four complex components, each one is written as a product of a module times a phase. In general, therefore, there would be four different modules as well as four different phases, all mixing between each other under Lorentz transformations \cite{Holland}.

Such a difficulty can be found ubiquitously in the literature, not only in the works about the polar form, but more in general in all investigations that aim at studying spinors in terms of tensor quantities. The idea of writing a spinor in terms of tensor quantities dates back to Cartan himself, and it continued immediately after him with the works of Whittaker \cite{W}, Ruse \cite{R} and Taub \cite{T}. In the same years Yvon \cite{Yvon1940} and later Takabayasi \cite{tr1,tr2} also contributed to this enterprise. And in more recent times new elements were brought in the game by Hestenes \cite{h1,h2,h3,h4,h5} in terms of the so-called space-time algebra. For a very comprehensive and clear account of all these results, and a more extensively detailed source of references, we address the interested reader to the very recent book of Zhelnorovich \cite{Book}.

Notwithstanding, there remains the mentioned issue of manifest covariance in general cases. While the cumulative results clearly support the idea that spinors, in spite of being complex objects, can nevertheless be written in terms of real tensors, these approaches are either tackling the problem component-by-component, thus lacking the manifest form of covariance \cite{W,R,T,Yvon1940}, or treating it in a manifest covariant way, although restricted to specially-relativistic situations \cite{tr1,tr2,h1,h2,h3,h4,h5}, or performing the study for generally-relativistic cases in curved space-times, but always in a preferred basis \cite{Book}. Neither in these works, nor anywhere else in the literature, a manifestly covariant general study is found.

In this quest for generality, the attempt that went farthest, for what we are aware, is that of Jakobi and Lochak in \cite{jl1, jl2}, where the spinor field was written in terms of real tensors by means of the polar decomposition without the use of any preferred basis, although again we have no knowledge of any attempt to investigate such polar decomposition at a differential level. For what we can tell, it is only very recently that, upon introduction of suitable objects called tensorial connections, it has become possible to write the polar form of the covariant derivative of the spinor field in such a way that its structure is manifestly covariant under general transformations \cite{Fabbri:2021mfc}.

And correspondingly, it has finally become possible to write the hydrodynamic form of the Dirac equations in such a way that its structure is manifestly covariant under general transformations, in space-times that are curved, with torsion, and in presence of electrodynamics \cite{Fabbri:2020ypd}. In this paper we will ask whether this hydrodynamic form is unique, and we will demonstrate that this is not so. We shall prove that there are in fact $19$ different reconfigurations of the hydrodynamic form of the Dirac equations, and that such $19$ systems of field equations are the minimal ones that are equivalent to one another, in the sense that from any one of such systems there is not a single field equation that can be removed without producing also the loss of equivalence to the original Dirac equations.
\section{Dirac Hydrodynamics}
\subsection{Kinematic Quantities}
\subsubsection{Spinor Fields in Polar Form}
We start this first section by recalling the general ideas of the conversion of the Dirac theory in its hydrodynamic formulation. As stated in the introduction, for us \emph{Dirac hydrodynamics}, or relativistic spinning quantum mechanics in hydrodynamic form, is just the relativistic extension with spin of quantum mechanics in hydrodynamic form as it was initially conceived by Madelung. As such, it will also be taken as synonym to \emph{spinor theory in polar form}.

We begin the presentation by recalling the notations. First of all, we define the Clifford matrices $\boldsymbol{\gamma}^{a}$ as the set of matrices verifying $\{\boldsymbol{\gamma}_{a},\boldsymbol{\gamma}_{b}\}\!=\!2\mathbb{I}\eta_{ab}$ where $\eta_{ab}$ is the Minkowski matrix. From them we define $[\boldsymbol{\gamma}_{a},\boldsymbol{\gamma}_{b}]/4\!=\!\boldsymbol{\sigma}_{ab}$ verifying $2i\boldsymbol{\sigma}_{ab}\!=\!\varepsilon_{abcd}\boldsymbol{\pi}\boldsymbol{\sigma}^{cd}$ where $\varepsilon_{abcd}$ is the completely antisymmetric pseudo-tensor and in which the $\boldsymbol{\pi}$ matrix\footnote{This matrix is what is traditionally indicated by $\boldsymbol{\gamma}^{5}$ as the fifth matrix after $\boldsymbol{\gamma}^{1}$, $\boldsymbol{\gamma}^{2}$, $\boldsymbol{\gamma}^{3}$, $\boldsymbol{\gamma}^{4}$ in Kaluza-Klein theories, at the time when the temporal coordinate was still designated as fourth coordinate. Nowadays we use a zero to indicate the temporal gamma matrix and thus we should use a four for the fifth gamma matrix, or better still we should just drop an index that no longer corresponds to any actual dimension. But even worse, in some signatures the position of the index five, whether upper or lower, changes sign of the gamma matrix, with ensuing risks of errors that would become possible. To avoid the potential for mistakes, we prefer to employ a notation in which there is no index at all. The choice of $\boldsymbol{\pi}$ instead of $\boldsymbol{\gamma}$ is because that matrix is parity-odd as opposed to the usual gammas being parity-even. This is not dissimilar to the normally accepted choice of using $\boldsymbol{\sigma}$ for the commutators of gamma matrices.} is implicitly defined. The identity $\boldsymbol{\gamma}_{i}\boldsymbol{\gamma}_{j}\boldsymbol{\gamma}_{k}\!=\! \boldsymbol{\gamma}_{i}\eta_{jk}-\boldsymbol{\gamma}_{j}\eta_{ik}+\boldsymbol{\gamma}_{k}\eta_{ij} +i\varepsilon_{ijkq}\boldsymbol{\pi}\boldsymbol{\gamma}^{q}$ shows that any product of more than two matrices can always be reduced to two matrices, therefore suggesting that the set $(\mathbb{I},\ \boldsymbol{\gamma}^{a},\ \boldsymbol{\sigma}_{ab},\ \boldsymbol{\gamma}^{a}\boldsymbol{\pi},\ \boldsymbol{\pi})$ is complete, and so it is also a basis for the Clifford space. From these definitions it is straightforward to see that $\boldsymbol{\sigma}_{ab}$ are the generators of the Lorentz algebra. Exponentiation of the generators gives $\boldsymbol{\Lambda}$ as the element of the Lorentz group. We notice that $\boldsymbol{\Lambda}$ is a complex Lorentz transformation, as opposed to $\Lambda^{a}_{b}$ being the real Lorentz transformation. The two are linked by the relation $\boldsymbol{\Lambda}\boldsymbol{\gamma}^{b}\boldsymbol{\Lambda}^{-1}\Lambda^{a}_{b}\!=\!\boldsymbol{\gamma}^{a}$ which in turns also specifies that all the Clifford matrices are constant, as is expected and well known. With the complex Lorentz transformation $\boldsymbol{\Lambda}$ and a general unitary phase $e^{iq\alpha}$ we define $\boldsymbol{S}\!=\!\boldsymbol{\Lambda}e^{iq\alpha}$ as the spinorial transformation. The spinorial transformation is therefore just the product of boosts and rotations as well as a gauge shift of charge $q$ as it is supposed to be to account for both space-time and electrodynamic transformations.

So far, nothing is new, and, aside for the specific convention we employ, all can be found in textbooks. With these tools, we define spinor fields as objects that, under spinorial transformations, transform according to
\begin{eqnarray}
&\psi\!\rightarrow\!\boldsymbol{S}\psi\ \ \ \ \ \ \ \ \mathrm{and} 
\ \ \ \ \ \ \ \ \ \overline{\psi}\!\rightarrow\!\overline{\psi}\boldsymbol{S}^{-1}
\end{eqnarray}
where $\overline{\psi}\!=\!\psi^{\dagger}\boldsymbol{\gamma}^{0}$ is the (unique) adjunction procedure. With these adjoint spinors, we can construct the bi-linears
\begin{eqnarray}
&\Sigma^{ab}\!=\!2\overline{\psi}\boldsymbol{\sigma}^{ab}\boldsymbol{\pi}\psi\ \ \ \ 
\ \ \ \ \ \ \ \ M^{ab}\!=\!2i\overline{\psi}\boldsymbol{\sigma}^{ab}\psi\\
&S^{a}\!=\!\overline{\psi}\boldsymbol{\gamma}^{a}\boldsymbol{\pi}\psi\ \ \ \ 
\ \ \ \ \ \ \ \ U^{a}\!=\!\overline{\psi}\boldsymbol{\gamma}^{a}\psi\\
&\Theta\!=\!i\overline{\psi}\boldsymbol{\pi}\psi\ \ \ \ 
\ \ \ \ \ \ \ \ \Phi\!=\!\overline{\psi}\psi\label{scalars}
\end{eqnarray}
which are all real tensors. We have defined six of them for reasons of symmetry and simplicity, but as it is clear from the linear independence of the Clifford matrices, they are not all linearly independent. In fact
\begin{eqnarray}
&\Sigma^{ij}\!=\!-\frac{1}{2}\varepsilon^{abij}M_{ab}
\end{eqnarray}
showing that the two antisymmetric tensors are just the Hodge dual of one another. By taking only one of them, $M_{ab}$ for example, together with the other four bi-linears, we may form a set of bi-linears that are all linearly independent, although not independent. Indeed
\begin{eqnarray}
&M_{ab}(\Phi^{2}\!+\!\Theta^{2})\!=\!\Phi U^{j}S^{k}\varepsilon_{jkab}\!+\!\Theta U_{[a}S_{b]}
\end{eqnarray}
showing that if $\Phi^{2}\!+\!\Theta^{2}\!\neq\!0$ then also $M_{ab}$ can be dropped in favour of the two vectors and the two scalars. Axial-vector and vector with pseudo-scalar and scalar are also not independent since
\begin{eqnarray}
&U_{a}U^{a}\!=\!-S_{a}S^{a}\!=\!\Theta^{2}\!+\!\Phi^{2}\label{norm}\\
&U_{a}S^{a}\!=\!0\label{orthogonal}
\end{eqnarray}
and in the case in which $\Phi^{2}\!+\!\Theta^{2}\!\neq\!0$ we can see that the axial-vector is space-like while the vector is time-like. Finally
\begin{eqnarray}
&2\boldsymbol{\sigma}^{\mu\nu}U_{\mu}S_{\nu}\boldsymbol{\pi}\psi\!+\!U^{2}\psi=0\label{a1}\\
&i\Theta S_{\mu}\boldsymbol{\gamma}^{\mu}\psi
\!+\!\Phi S_{\mu}\boldsymbol{\gamma}^{\mu}\boldsymbol{\pi}\psi\!+\!U^{2}\psi=0\label{a2}
\end{eqnarray}
to complete the list of bi-linear identities. The list can be longer, but we will not need any more than these.

Now, a result that can be proven is that under the general assumption $\Phi^{2}\!+\!\Theta^{2}\!\neq\!0$ it is always possible to write any spinor field in polar form, which in chiral representation is given according to
\begin{eqnarray}
&\psi\!=\!\phi\ e^{-\frac{i}{2}\beta\boldsymbol{\pi}}
\ \boldsymbol{L}^{-1}\left(\begin{tabular}{c}
$1$\\
$0$\\
$1$\\
$0$
\end{tabular}\right)
\label{spinor}
\end{eqnarray}
for a pair of functions $\phi$ and $\beta$ and for some $\boldsymbol{L}$ having the structure of a spinorial transformation and that such polar form is unique up to the discrete transformation $\beta\!\rightarrow\!\beta\!+\!\pi$ and up to the reversal of the third axis \cite{jl1, jl2}. The three elements $\boldsymbol{L}$, $\phi$ and $\beta$ in \eqref{spinor} need some explanation. After that \eqref{spinor} is substituted in (\ref{scalars}) one gets
\begin{eqnarray}
&\Theta\!=\!2\phi^{2}\sin{\beta}\ \ \ \ 
\ \ \ \ \ \ \ \ \Phi\!=\!2\phi^{2}\cos{\beta}
\end{eqnarray}
showing that $\phi$ and $\beta$ are a real scalar and a real pseudo-scalar, called module and chiral angle. Being equipped with the polar form we can also normalize
\begin{eqnarray}
&S^{a}\!=\!2\phi^{2}s^{a}\ \ \ \ 
\ \ \ \ \ \ \ \ U^{a}\!=\!2\phi^{2}u^{a}
\end{eqnarray}
where $u^{a}$ and $s^{a}$ are the velocity vector and the spin axial-vector. The identities (\ref{norm}-\ref{orthogonal}) reduce to
\begin{eqnarray}
&u_{a}u^{a}\!=\!-s_{a}s^{a}\!=\!1\\
&u_{a}s^{a}\!=\!0
\end{eqnarray}
showing that the velocity has only $3$ independent components, given by the $3$ components of its spatial part (indeed the constraint $u_{a}u^{a}\!=\!1$ fixes the temporal component), whereas the spin has only $2$ independent components, assigned by the $2$ angles that, in the rest-frame, its spatial part forms with the third axis (in fact because of the constraint $u_{a}s^{a}\!=\!0$ the temporal component of the spin is zero in the rest-frame, and due to the constraint $s_{a}s^{a}\!=\!-1$ one of the spatial components is fixed). The physical interpretation of this mathematical fact is that, in the frame at rest, there remains no spatial component of the velocity and its temporal component is unity, and because this also implies the vanishing of the temporal component of the spin then its spatial component, when aligned along the third axis, selects this axis as the axis of symmetry of the system. Hence, any rotation around this axis would have to be an irrelevant rotation, as it would be unable to have any effect on the spinor itself. In the following, we will still refer to rotations around the third axis because we have chosen this axis for all explicit computations, although it is clear that from a covariant perspective we would refer to them as rotations around the spin axis. The remaining identities (\ref{a1}-\ref{a2}) are
\begin{eqnarray}
&2\boldsymbol{\sigma}^{\mu\nu}u_{\mu}s_{\nu}\boldsymbol{\pi}\psi\!+\!\psi=0\label{aux1}\\
&is_{\mu}\boldsymbol{\gamma}^{\mu}\psi\sin{\beta}
\!+\!s_{\mu}\boldsymbol{\gamma}^{\mu}\boldsymbol{\pi}\psi\cos{\beta}\!+\!\psi=0\label{aux2}
\end{eqnarray}
which will be useful later on. As for $\boldsymbol{L}$ we know that it has the structure of a spinorial transformation, and thus it is a product of a gauge times a Lorentz transformation. Its interpretation is of straightforward reading. It is the specific transformation that takes an assigned spinor into its rest frame with spin aligned along the third axis, however general the initial spinor is. As such $\boldsymbol{L}$ should not be confused with $\boldsymbol{S}$ because even if mathematically they are the same, and that is they are both a spinorial transformation, from a physical perspective they are very different, with $\boldsymbol{S}$ telling in what way spinors behave under transformations and $\boldsymbol{L}$ telling how any given spinor can always be seen as a suitable deformation of its simplest rest-frame spin-eigenstate form. Metaphorically, if the spinor were to be a top spinning on a table then $\boldsymbol{S}$ would tell how to move from the fixed system of reference in which the table is at rest to the rotating system of reference in which the top is at rest while $\boldsymbol{L}$ would tell how the top is spinning. So the advantage of writing spinor fields in polar form is that in their $4$ complex components, the $8$ real functions are re-organized in such a way that the $2$ real scalars, that is the two true degrees of freedom ($\phi$ and $\beta$), remain isolated from the $6$ real parameters, which can always be transferred into the frame (encoded within $\boldsymbol{L}$). The fact that $\boldsymbol{L}$ is the product of a gauge times a Lorentz transformation might in principle lead us to think that it has in total $1\!+\!6\!=\!7$ parameters and this appears to be in contradiction with the fact that it effectively has only $6$ parameters as we have just explained. The resolution of this paradox is that, as is clear from (\ref{spinor}), the action of a gauge transformation is indistinguishable from the action of a rotation around the third axis. The consequence is that there is a redundancy between the phase and the angle of rotation around the third axis, so that the a priori $7$ parameters are in fact reduced to $6$ parameters alone. If the total parameters of the combined gauge and Lorentz transformations are essentially $6$ and $1$ is encoded by the gauge transformation then there can be no more than $5$ parameters that remain in the Lorentz transformation, and this is precisely what happens. The $5$ parameters are given by the $3$ rapidities of the velocity vector and the $2$ Euler angles of the spin axial-vector, as we have discussed above. This redundancy should not be surprising, and indeed it is rather common. An identical situation happens in the Standard Model, where the $\mathrm{U(1)\!\times\!SU(2)}$ group has $4$ parameters, but the hypercharge and the third component of the isospin combine to form a single parameter, so that the entire group does not have $4$ but $3$ Goldstone bosons solely. We notice that this is more than an example, it is a true mathematical analogy. Indeed, the fact that the parameters of $\boldsymbol{L}$ can always be transferred into the frame means that they are the Goldstone bosons associated to the spinor field by definition. The Goldstone bosons we have here play for the spinor field exactly the same role that the Goldstone bosons in the Standard Model play for the Higgs field as demonstrated in reference \cite{Fabbri:2021mfc}. We will be back to this after discussing what happens at the differential level.

For now, we notice that the preferred frame mentioned in \cite{Book} is the frame in which $\boldsymbol{L}\!=\!\mathbb{I}$ with the spinor at rest and in spin eigenstate. The generality achieved in \cite{jl1, jl2} is hence due to the $\boldsymbol{L}$ operator.
\subsubsection{Tensorial Connections}
For general spinor fields the spinorial covariant derivative is defined according to
\begin{eqnarray}
&\boldsymbol{\nabla}_{\mu}\psi\!=\!\partial_{\mu}\psi
\!+\!\boldsymbol{C}_{\mu}\psi\label{spincovder}
\end{eqnarray}
in terms of the spinorial connection $\boldsymbol{C}_{\mu}$ which is itself defined by its transformation law
\begin{eqnarray}
&\boldsymbol{C}_{\mu}\!\rightarrow\!\boldsymbol{S}\left(\boldsymbol{C}_{\mu}
\!-\!\boldsymbol{S}^{-1}\partial_{\mu}\boldsymbol{S}\right)\boldsymbol{S}^{-1}
\label{spinconn}
\end{eqnarray}
where $\boldsymbol{S}$ is the spinorial transformation law. This spinorial connection can be decomposed according to
\begin{eqnarray}
&\boldsymbol{C}_{\mu}\!=\!\frac{1}{2}C^{ab}_{\phantom{ab}\mu}\boldsymbol{\sigma}_{ab}
\!+\!iqA_{\mu}\boldsymbol{\mathbb{I}}\label{spinorialconnection}
\end{eqnarray}
where $C^{ab}_{\phantom{ab}\mu}$ is the spin connection of the tetrads of the space-time and $A_{\mu}$ is the gauge potential. We will not give in the present work the details of the explicit form of the spin connection written in terms of the tetrad fields because this is very standard material of any introductory course of General Relativity in the tetradic formalism \cite{Book}.

Now, when in the spinorial covariant derivative the spinor field is written in polar form, something interesting starts to emerge. To see what, we recall that one can demonstrate that it is possible to write
\begin{eqnarray}
&\boldsymbol{L}^{-1}\partial_{\mu}\boldsymbol{L}\!=\!iq\partial_{\mu}\xi\mathbb{I}
\!+\!\frac{1}{2}\partial_{\mu}\xi_{ij}\boldsymbol{\sigma}^{ij}\label{spintrans}
\end{eqnarray}
for some $\partial_{\mu}\xi$ and $\partial_{\mu}\xi_{ij}$ which are in fact the Goldstone fields of the spinor mentioned above. In \eqref{spintrans} the redundancy between phase and angle of rotation around the third axis shows that the apparent $7$ Goldstone fields are effectively $6$ Goldstone fields only, since $\partial_{\mu}\xi_{12}$ can always be re-absorbed in a redefinition of $q\partial_{\mu}\xi$ (this can be seen considering that their respective generators $\boldsymbol{\sigma}^{12}$ and $\mathbb{I}$ act identically on the rest-frame spin-eigenstate spinor by construction). Also, in $q\partial_{\mu}\xi$ the presence of the charge has been made explicit to render simpler the definition of the following quantities
\begin{eqnarray}
&q(\partial_{\mu}\xi\!-\!A_{\mu})\!\equiv\!P_{\mu}\label{P}\\
&\partial_{\mu}\xi_{ij}\!-\!C_{ij\mu}\!\equiv\!R_{ij\mu}\label{R}
\end{eqnarray}
since now $q$ can be collected in the definition of $P_{\mu}$ in (\ref{P}). It is now possible to see that, after that the Goldstone fields are transferred into the frame, they combine with gauge potential and spin connection so to become the longitudinal components of the $P_{\mu}$ and $R_{ij\mu}$ objects, which then turn into a real vector and a real tensor called gauge and space-time tensorial connections, as it has been demonstrated in \cite{Fabbri:2021mfc}. With (\ref{P}-\ref{R}) we finally have that
\begin{eqnarray}
&\!\!\!\!\!\!\!\!\boldsymbol{\nabla}_{\mu}\psi\!=\!(-\frac{i}{2}\nabla_{\mu}\beta\boldsymbol{\pi}
\!+\!\nabla_{\mu}\ln{\phi}\mathbb{I}
\!-\!iP_{\mu}\mathbb{I}\!-\!\frac{1}{2}R_{ij\mu}\boldsymbol{\sigma}^{ij})\psi
\label{decspinder}
\end{eqnarray}
as the polar form of the spinor field covariant derivative. From it we can deduce that
\begin{eqnarray}
&\nabla_{\mu}s_{i}\!=\!R_{ji\mu}s^{j}\ \ \ \
\ \ \ \ \ \ \ \ \nabla_{\mu}u_{i}\!=\!R_{ji\mu}u^{j}\label{ds,du}
\end{eqnarray}
are valid as general identities. Recall that rotations around the spin axis are unable to have effects on any component of $s_{i}$ and $u_{i}$ and therefore on $\nabla_{\nu}s_{i}$ and $\nabla_{\nu}u_{i}$ in general. Because of (\ref{ds,du}) this means that the components of $R_{ab\nu}$ that correspond to the rotations around the spin axis will remain undetermined. From (\ref{ds,du}) we get the identities
\begin{eqnarray}
&R_{ab\mu}\!\equiv\!u_{a}\nabla_{\mu}u_{b}\!-\!u_{b}\nabla_{\mu}u_{a}
\!+\!s_{b}\nabla_{\mu}s_{a}\!-\!s_{a}\nabla_{\mu}s_{b}
\!+\!(u_{a}s_{b}\!-\!u_{b}s_{a})\nabla_{\mu}u_{k}s^{k}
\!+\!\frac{1}{2}R_{ij\mu}\varepsilon^{ijcd}\varepsilon_{abpq}s_{c}u_{d}s^{p}u^{q}
\end{eqnarray}
so that we can write
\begin{eqnarray}
&R_{ab\mu}\!=\!u_{a}\nabla_{\mu}u_{b}\!-\!u_{b}\nabla_{\mu}u_{a}
\!+\!s_{b}\nabla_{\mu}s_{a}\!-\!s_{a}\nabla_{\mu}s_{b}
\!+\!(u_{a}s_{b}\!-\!u_{b}s_{a})\nabla_{\mu}u_{k}s^{k}
\!+\!2\varepsilon_{abij}u^{i}s^{j}V_{\mu}\label{Rfull}
\end{eqnarray}
for some vector $V_{\mu}\!=\!\frac{1}{4}R_{ij\mu}\varepsilon^{ijcd}u_{c}s_{d}$ that is no more specified. This vector is precisely what represents the components of $R_{ab\nu}$ that correspond to rotations around the spin axis, as clear from the fact that $2V_{\mu}\!=\!R_{12\mu}$ whenever the spinor field is in its rest frame with spin aligned along the third axis. As we have commented in the previous sub-section, a general spinor field can always be seen as the simplest spinor field at rest and having spin aligned along the third axis after a suitable deformation, and as we see now, such a deformation $\boldsymbol{L}$ has the structure of a spinorial transformation in which the generator $ij$ has a parameter whose derivative with respect to the $\mu$-th coordinate is given by the component $R_{ij\mu}$ of the tensorial connection. As the deformation $\boldsymbol{L}$ can be interpreted like a strain, the tensorial connection $R_{ij\mu}$ is interpretable like the strain-rate tensor. This analogy is very strong if we consider that $R_{ij\mu}$ in \eqref{Rfull} can be written in terms of derivatives of spin and velocity, and that the derivative of the velocity is the same object with which the strain-rate tensor is constructed in continuum mechanics. Together, gauge and space-time tensorial connections $R_{ij\mu}$ and $P_{\mu}$ have a total of $24+4\!=\!28$ components, so that once we subtract the $4$ components that cannot be determined, we remain with an effective total of $24$ components. These $24$ components, added to the $4\!\times\!2\!=\!8$ components of $\nabla_{\mu}\phi$ and $\nabla_{\mu}\beta$ make up for the full $32$ components of the spinor field covariant derivative. The full counting of components is perfectly matched. As for the parallel with the Standard Model that we have discussed in the previous sub-section, we can say that the tensorial connections $R_{ij\mu}$ and $P_{\mu}$ are just the geometric and electrodynamic analog of the vector bosons $W_{\nu}^{\pm}$ and $Z_{\nu}$ and the frame in which the spinor field is at rest and with spin along the third axis is the analog of the unitary gauge \cite{Fabbri:2021mfc}. As we said, these analogies are due to a strict mathematical parallel between the polar form of spinor fields and the Higgs field in the Standard Model, a parallel holding up until symmetry breaking.

To complete the comparison with previous works, we can finally say that it is precisely the definition of the tensorial connection what allowed the results of Jakobi and Lochak \cite{jl1, jl2} on the polar form of the spinor field to be extended to its differential structures \cite{Fabbri:2021mfc}. And from this, we can now move to the dynamics.
\subsection{Dynamical Equations}
In the previous sub-section we have introduced the differential structures, that is the gauge and space-time tensorial connections, with which to define the covariant derivative of spinor fields in polar form. Readers may have noticed that everything was done by taking into account the tetradic structure, and therefore the metric structure, of the space-time itself. Nevertheless, we have neglected all torsional degrees of freedom. A reader that cares about differential geometry in its most general form might now complain that torsion should be allowed instead. In order for this to be done one can simply notice that, because torsion is a true tensor, it is always possible to decompose the most general connection into the torsionless connection plus the torsional contributions. Therefore, the most general differential geometry with torsion is always equivalent to the differential geometry with no torsion with an additional field representing the torsion tensor. As a consequence, full generality will be restored in the dynamics by the addition of the torsion field. Because torsion couples to spin, which in the case of Dirac spinors is completely antisymmetric, torsion will have a completely antisymmetric part only, which is equivalent to the Hodge dual of an axial-vector. Hence, in the dynamics, we shall add torsion in the form of an axial-vector field. The reader interested in more details can find them in \cite{Fabbri:2017lmf}.

So, the dynamics of the spinor field will be given here in terms of the Dirac equations
\begin{eqnarray}
&i\boldsymbol{\gamma}^{\mu}\boldsymbol{\nabla}_{\mu}\psi
\!-\!XW_{\mu}\boldsymbol{\gamma}^{\mu}\boldsymbol{\pi}\psi\!-\!m\psi\!=\!0
\label{D}
\end{eqnarray}
with $W_{i}$ the axial-vector torsion and $X$ the coupling constant of the torsion-spin interaction. Now by multiplying (\ref{D}) on the left with $\mathbb{I}$, or $\boldsymbol{\gamma}^{a}$, or $\boldsymbol{\sigma}_{ab}$, or $\boldsymbol{\gamma}^{a}\boldsymbol{\pi}$, or $\boldsymbol{\pi}$, and then, in each case, also with $\overline{\psi}$, we obtain five equations that, after splitting real and imaginary parts, give ten equations that are all real and tensorial in structure. They are given by
\begin{eqnarray}
&\nabla_{\mu}U^{\mu}\!=\!0\label{divU}\\
&\frac{i}{2}(\overline{\psi}\boldsymbol{\gamma}^{\mu}\boldsymbol{\pi}\boldsymbol{\nabla}_{\mu}\psi
\!-\!\boldsymbol{\nabla}_{\mu}\overline{\psi}\boldsymbol{\gamma}^{\mu}\boldsymbol{\pi}\psi)
\!-\!XW_{\sigma}U^{\sigma}\!=\!0\label{Lodd}\\
&\nabla^{[\alpha}U^{\nu]}\!+\!i\varepsilon^{\alpha\nu\mu\rho}
(\overline{\psi}\boldsymbol{\gamma}_{\rho}\boldsymbol{\pi}\!\boldsymbol{\nabla}_{\mu}\psi\!-\!\!
\boldsymbol{\nabla}_{\mu}\overline{\psi}\boldsymbol{\gamma}_{\rho}\boldsymbol{\pi}\psi)
\!-\!2XW_{\sigma}U_{\rho}\varepsilon^{\alpha\nu\sigma\rho}\!-\!2mM^{\alpha\nu}\!=\!0\label{curlU}
\end{eqnarray}
\begin{eqnarray}
&\nabla_{\mu}S^{\mu}\!-\!2m\Theta\!=\!0\label{divS}\\
&\frac{i}{2}(\overline{\psi}\boldsymbol{\gamma}^{\mu}\boldsymbol{\nabla}_{\mu}\psi
\!-\!\boldsymbol{\nabla}_{\mu}\overline{\psi}\boldsymbol{\gamma}^{\mu}\psi)
\!-\!XW_{\sigma}S^{\sigma}\!-\!m\Phi\!=\!0\label{Leven}\\
&\nabla^{\mu}S^{\rho}\varepsilon_{\mu\rho\alpha\nu}
\!+\!i(\overline{\psi}\boldsymbol{\gamma}_{[\alpha}\!\boldsymbol{\nabla}_{\nu]}\psi
\!-\!\!\boldsymbol{\nabla}_{[\nu}\overline{\psi}\boldsymbol{\gamma}_{\alpha]}\psi)
\!+\!2XW_{[\alpha}S_{\nu]}\!=\!0\label{curlS}
\end{eqnarray}
\begin{eqnarray}
&i(\overline{\psi}\boldsymbol{\nabla}^{\alpha}\psi
\!-\!\boldsymbol{\nabla}^{\alpha}\overline{\psi}\psi)
\!-\!\nabla_{\mu}M^{\mu\alpha}
\!-\!XW_{\sigma}M_{\mu\nu}\varepsilon^{\mu\nu\sigma\alpha}\!-\!2mU^{\alpha}\!=\!0
\label{vr}\\
&(\boldsymbol{\nabla}_{\alpha}\overline{\psi}\boldsymbol{\pi}\psi
\!-\!\overline{\psi}\boldsymbol{\pi}\boldsymbol{\nabla}_{\alpha}\psi)
\!-\!\frac{1}{2}\nabla^{\mu}M^{\rho\sigma}\varepsilon_{\rho\sigma\mu\alpha}
\!+\!2XW^{\mu}M_{\mu\alpha}\!=\!0\label{ai}
\end{eqnarray}
\begin{eqnarray}
&\nabla_{\alpha}\Phi
\!-\!2(\overline{\psi}\boldsymbol{\sigma}_{\mu\alpha}\!\boldsymbol{\nabla}^{\mu}\psi
\!-\!\boldsymbol{\nabla}^{\mu}\overline{\psi}\boldsymbol{\sigma}_{\mu\alpha}\psi)
\!+\!2X\Theta W_{\alpha}\!=\!0\label{vi}\\
&\nabla_{\nu}\Theta\!-\!
2i(\overline{\psi}\boldsymbol{\sigma}_{\mu\nu}\boldsymbol{\pi}\boldsymbol{\nabla}^{\mu}\psi\!-\!
\boldsymbol{\nabla}^{\mu}\overline{\psi}\boldsymbol{\sigma}_{\mu\nu}\boldsymbol{\pi}\psi)
\!-\!2X\Phi W_{\nu}\!+\!2mS_{\nu}\!=\!0\label{ar}
\end{eqnarray}
which are the well-known Gordon decompositions of the Dirac equations.

Plugging into all of them the polar form of the spinor field and its covariant derivative allows us to obtain
\begin{eqnarray}
&\nabla_{\mu}U^{\mu}\!=\!0\label{poldivU}\\
&(\nabla_{\mu}\beta\!-\!2XW_{\mu}
\!+\!\frac{1}{2}\varepsilon_{\mu\alpha\nu\rho}R^{\alpha\nu\rho})U^{\mu}
\!+\!2P_{\mu}S^{\mu}\!=\!0\label{polLodd}\\
&\nabla^{[\alpha}U^{\nu]}
\!+\!\varepsilon^{\alpha\nu\mu\rho}(\nabla_{\mu}\beta\!-\!2XW_{\mu})U_{\rho}
\!-\!\frac{1}{2}R^{ij}_{\phantom{ij}\mu}\varepsilon_{ij\rho\kappa}
U^{\kappa}\varepsilon^{\alpha\nu\mu\rho}
\!+\!2\varepsilon^{\alpha\nu\mu\rho}P_{\mu}S_{\rho}\!-\!2mM^{\alpha\nu}\!=\!0\label{polcurlU}
\end{eqnarray}
\begin{eqnarray}
&\nabla_{\mu}S^{\mu}\!-\!2m\Theta\!=\!0\label{poldivS}\\
&(\nabla_{\mu}\beta\!-\!2XW_{\mu}
\!+\!\frac{1}{2}\varepsilon_{\mu\alpha\nu\rho}R^{\alpha\nu\rho})S^{\mu}
\!+\!2P_{\mu}U^{\mu}\!-\!2m\Phi\!=\!0\label{polLeven}\\
&\nabla^{[\alpha}S^{\nu]}
\!+\!\varepsilon^{\alpha\nu\mu\rho}(\nabla_{\mu}\beta\!-\!2XW_{\mu})S_{\rho}
\!-\!\frac{1}{2}R^{ij}_{\phantom{ij}\mu}\varepsilon_{ij\rho\kappa}
S^{\kappa}\varepsilon^{\alpha\nu\mu\rho}
\!+\!2\varepsilon^{\alpha\nu\mu\rho}P_{\mu}U_{\rho}\!=\!0\label{polcurlS}
\end{eqnarray}
\begin{eqnarray}
&\!\!\!\!\!\!\!\!\nabla_{\mu}M^{\mu\alpha}\!-\!2XW_{\sigma}\Sigma^{\sigma\alpha}
\!+\!\frac{1}{2}R^{ij\alpha}M_{ij}\!-\!2P^{\alpha}\Phi
\!+\!2mU^{\alpha}\!=\!0\label{polvr}\\
&\!\!\!\!\nabla^{\mu}\Sigma_{\mu\alpha}\!+\!2XW^{\mu}M_{\mu\alpha}
\!+\!\frac{1}{2}R_{ij\alpha}\Sigma^{ij}\!+\!2P_{\alpha}\Theta\!=\!0\label{polai}
\end{eqnarray}
\begin{eqnarray}
&\nabla_{\alpha}\Phi\!+\!(2XW_{\alpha}
\!-\!\frac{1}{2}\varepsilon_{\alpha\mu\nu\rho}R^{\mu\nu\rho})\Theta
\!+\!R_{\alpha\sigma}^{\phantom{\alpha\sigma}\sigma}\Phi\!+\!2P^{\mu}M_{\mu\alpha}\!=\!0\label{polvi}\\
&\nabla_{\nu}\Theta\!-\!(2XW_{\nu}
\!-\!\frac{1}{2}\varepsilon_{\nu\mu\sigma\rho}R^{\mu\sigma\rho})\Phi
\!+\!R_{\nu\sigma}^{\phantom{\nu\sigma}\sigma}\Theta\!-\!2P^{\mu}\Sigma_{\mu\nu}
\!+\!2mS_{\nu}\!=\!0\label{polar}
\end{eqnarray}
which are the Gordon decompositions in polar form. The last two, after a proper diagonalization, give
\begin{eqnarray}
&\nabla_{\mu}\beta\!-\!2XW_{\mu}\!+\!B_{\mu}
\!-\!2P^{\iota}u_{[\iota}s_{\mu]}
\!+\!2ms_{\mu}\cos{\beta}\!=\!0\label{b}\\
&\nabla_{\mu}\ln{\phi^{2}}\!+\!R_{\mu}
\!-\!2P^{\rho}u^{\nu}s^{\alpha}\varepsilon_{\mu\rho\nu\alpha}
\!+\!2ms_{\mu}\sin{\beta}\!=\!0\label{m}
\end{eqnarray}
specifying all derivatives of both degrees of freedom in terms of $R_{\nu a}^{\phantom{\nu a}a}\!=\!R_{\nu}$ and $\varepsilon_{\nu\alpha\pi\iota}R^{\alpha\pi\iota}/2\!=\!B_{\nu}$ as well as $P_{\mu}$ and of course the torsion of the space-time. By means of the identities (\ref{aux1}-\ref{aux2}) one can prove that these two equations imply the Dirac equation \eqref{D} in polar form, and thus in general. Hence, (\ref{b}-\ref{m}) are equivalent to the Dirac equation \cite{Fabbri:2020ypd}.

The equations (\ref{b}-\ref{m}) can then be called Dirac equations in polar form. They express the Dirac dynamics by means of the degrees of freedom $\phi^{2}$ and $\beta$ supplemented by $u_{\alpha}$ and $s_{\alpha}$ all of which having a clear meaning. Specifically, the chiral angle $\beta$ is the phase difference between chiral parts, $s_{\alpha}$ is the spin, the module squared $\phi^{2}$ is the density, $u_{\alpha}$ is the velocity of the matter distribution described by the spinor field. Necessary condition for the non-relativistic limit is to have $\beta\!=\!0$ while the classical approximation is implemented by asking $s_{\alpha}\!\rightarrow\!0$ \cite{Fabbri:2020ypd}. As for the density $\phi^{2}$ and the velocity $u_{\alpha}$ they are exactly the density and velocity one would have in fluid mechanics. Thus, equations (\ref{b}-\ref{m}) are what expresses the Dirac dynamics as a type of hydrodynamics, therefore extending to the relativistic case with spin the hydrodynamic formulation of quantum mechanics that was originally given by Madelung.
\section{19 Formulations}
We have seen that the Dirac theory can be written in hydrodynamic formulation. Our goal now is to find all ways in which this can be done, that is find all manners to write a covariant set of field equations like (\ref{b}-\ref{m}).

To make things easier, we start by introducing the two vectors
\begin{eqnarray}
&2E_{\mu}\!=\!B_{\mu}\!-\!2XW_{\mu}\!+\!\nabla_{\mu}\beta\!+\!2ms_{\mu}\cos{\beta}\label{E}\\
&2F_{\mu}\!=\!R_{\mu}\!+\!\nabla_{\mu}\ln{\phi^{2}}\!+\!2ms_{\mu}\sin{\beta}\label{F}
\end{eqnarray}
which, admittedly, is a definition that may appear arbitrary now, but which is essential to simplify all computations later. In fact, when in the full set of Gordon decompositions (\ref{poldivU}-\ref{polar}) we also substitute the bi-linear spinors in polar form, using the above definitions, we obtain, respectively, the following sets
\begin{eqnarray}
&F_{\mu}u^{\mu}\!=\!0\label{A1}\\
&E_{\mu}u^{\mu}\!+\!P_{\mu}s^{\mu}\!=\!0\label{A2}\\
&\varepsilon^{\alpha\nu\mu\rho}E_{\mu}u_{\rho}\!+\!F^{[\alpha}u^{\nu]}
\!+\!\varepsilon^{\alpha\nu\mu\rho}P_{\mu}s_{\rho}\!=\!0\label{A3}
\end{eqnarray}
\begin{eqnarray}
&F_{\mu}s^{\mu}\!=\!0\label{B1}\\
&E_{\mu}s^{\mu}\!+\!P_{\mu}u^{\mu}\!=\!0\label{B2}\\
&\varepsilon^{\alpha\nu\mu\rho}E_{\mu}s_{\rho}\!+\!F^{[\alpha}s^{\nu]}
\!+\!\varepsilon^{\alpha\nu\mu\rho}P_{\mu}u_{\rho}\!=\!0\label{B3}
\end{eqnarray}
\begin{eqnarray}
&F_{\mu}u_{j}s_{k}\varepsilon^{jk\mu\alpha}\!+\!E_{\mu}u^{[\mu}s^{\alpha]}\!-\!P^{\alpha}\!=\!0
\label{forwardmomentum}\\
&F_{\mu}u^{[\mu}s^{\alpha]}\!-\!E_{\mu}u_{j}s_{k}\varepsilon^{jk\mu\alpha}\!=\!0
\label{complmomentum}
\end{eqnarray}
\begin{eqnarray}
&F_{\mu}\!-\!P^{\rho}u^{\nu}s^{\alpha}\varepsilon_{\mu\rho\nu\alpha}\!=\!0\label{auxF}\\
&E_{\mu}\!-\!P^{\iota}u_{[\iota}s_{\mu]}\!=\!0\label{auxE}
\end{eqnarray}
as four groups of hydrodynamic field equations. The last group is of course (\ref{b}-\ref{m}) which we have already demonstrated to be equivalent to the Dirac equation. Therefore, any group that is proven to be equivalent to (\ref{auxF}-\ref{auxE}) will be equivalent to the Dirac equation. In fact, to be more precise, since the Dirac equation is already implying all groups, all we need to do is to prove that a group implies (\ref{auxF}-\ref{auxE}) to show the equivalence. This can be easily done with a formal tensor algebra. For instance, take (\ref{forwardmomentum}-\ref{complmomentum}), and then from \eqref{forwardmomentum} isolate $P^{\alpha}$ while from \eqref{complmomentum} project along $u^{\alpha}$ and $s^{\alpha}$ getting
\begin{eqnarray}
&P^{\alpha}\!=\!F_{i}u_{j}s_{k}\varepsilon^{ijk\alpha}\!+\!E_{i}u^{[i}s^{\alpha]}\label{momentum}\\
&E_{\mu}u_{j}s_{k}\varepsilon^{jk\mu\alpha}\!=\!0\label{constrE}\\
&F_{\mu}u^{\mu}\!=\!0\label{constrF1}\\
&F_{\mu}s^{\mu}\!=\!0\label{constrF2}
\end{eqnarray}
to be used in (\ref{auxF}-\ref{auxE}). By substituting \eqref{momentum} and repeatedly using \eqref{constrE} and (\ref{constrF1}-\ref{constrF2}), one can see that in fact (\ref{auxF}-\ref{auxE}) are verified. So (\ref{forwardmomentum}-\ref{complmomentum}) does imply (\ref{auxF}-\ref{auxE}). A similar proof can be done also for the first two groups. As a consequence, the groups (\ref{A1}-\ref{A2}-\ref{A3}), (\ref{B1}-\ref{B2}-\ref{B3}), (\ref{forwardmomentum}-\ref{complmomentum}), (\ref{auxF}-\ref{auxE}) are equivalent to one another, and each to the Dirac equations, as it has already been discussed in \cite{Fabbri:2023onb}. These are already $4$ ways in which the Dirac equations are written in hydrodynamic form. Every one, consisting of exactly $8$ equations, is also the most stringent to be equivalent to the Dirac equation.

Nevertheless, we may drop the strict equivalence to find other sets of hydrodynamic forms equivalent to the Dirac equation up to redundancies. To do this, we may choose the unitary gauge $u^{0}\!=\!1$ and $s^{3}\!=\!1$ getting, respectively for all groups, and in order within a single group, the following equations
\begin{eqnarray}
&F_{0}\!=\!0\label{F0}\\
&E_{0}\!+\!P_{3}\!=\!0\label{E0}\\
&F_{1}\!+\!P_{2}\!=\!0\ \ \ \ F_{2}\!-\!P_{1}\!=\!0\ \ \ \ F_{3}\!=\!0\ \ \ \ 
E_{3}\!+\!P_{0}\!=\!0\ \ \ \ E_{2}\!=\!0\ \ \ \ E_{1}\!=\!0\label{F3E3}
\end{eqnarray}
\begin{eqnarray}
&F_{3}\!=\!0\label{F3}\\
&E_{3}\!+\!P_{0}\!=\!0\label{E3}\\
&E_{2}\!=\!0\ \ \ \ E_{1}\!=\!0\ \ \ \ F_{0}\!=\!0\ \ \ \ 
E_{0}\!+\!P_{3}\!=\!0\ \ \ \ F_{1}\!+\!P_{2}\!=\!0\ \ \ \ F_{2}\!-\!P_{1}\!=\!0\label{F0E0}
\end{eqnarray}
\begin{eqnarray}
&E_{3}\!+\!P_{0}\!=\!0\ \ \ \ F_{2}\!-\!P_{1}\!=\!0\ \ \ \ 
F_{1}\!+\!P_{2}\!=\!0\ \ \ \ E_{0}\!+\!P_{3}\!=\!0\label{E0E3}\\
&F_{3}\!=\!0\ \ \ \ E_{2}\!=\!0\ \ \ \ 
E_{1}\!=\!0\ \ \ \ F_{0}\!=\!0\label{F0F3}
\end{eqnarray}
\begin{eqnarray}
&F_{0}\!=\!0\ \ \ \ F_{1}\!+\!P_{2}\!=\!0\ \ \ \ 
F_{2}\!-\!P_{1}\!=\!0\ \ \ \ F_{3}\!=\!0\label{Fi}\\
&E_{0}\!+\!P_{3}\!=\!0\ \ \ \ E_{1}\!=\!0\ \ \ \ 
E_{2}\!=\!0\ \ \ \ E_{3}\!+\!P_{0}\!=\!0\label{Ei}
\end{eqnarray}
and as is clear after some scrambling, all groups contain the same equations. Any grouping that would re-cover these $8$ equations would also obtain the validity of the Dirac equation in the unitary gauge, and so in general. For example, consider (\ref{A1}, \ref{B1}, \ref{forwardmomentum}, \ref{auxE}). They respectively yield (\ref{F0}, \ref{F3}, \ref{E0E3}, \ref{Ei}), which re-cover all equations, with $E_{0}\!+\!P_{3}\!=\!0$ and $E_{3}\!+\!P_{0}\!=\!0$ repeated twice. Hence the set (\ref{A1}, \ref{B1}, \ref{forwardmomentum}, \ref{auxE}) is equivalent to the Dirac equations, with $2$ redundant field equations. The same can be said for the set (\ref{A2}, \ref{B2}, \ref{complmomentum}, \ref{auxF}). After full inventory, we have the following sets:
\begin{enumerate}
\item (\ref{A1}, \ref{B1}, \ref{forwardmomentum}, \ref{auxE}), (\ref{A2}, \ref{B2}, \ref{complmomentum}, \ref{auxF}) with $2$
\item (\ref{A1}, \ref{A3}, \ref{forwardmomentum}), (\ref{A2}, \ref{A3}, \ref{complmomentum}), (\ref{A2}, \ref{A3}, \ref{auxF}), (\ref{A1}, \ref{A3}, \ref{auxE}), (\ref{B1}, \ref{B3}, \ref{forwardmomentum}), (\ref{B2}, \ref{B3}, \ref{complmomentum}), (\ref{B2}, \ref{B3}, \ref{auxF}), (\ref{B1}, \ref{B3}, \ref{auxE}) with $3$
\item (\ref{A3}, \ref{B3}) with $4$
\item (\ref{A3}, \ref{forwardmomentum}, \ref{auxF}), (\ref{A3}, \ref{complmomentum}, \ref{auxE}), (\ref{B3}, \ref{forwardmomentum}, \ref{auxF}), (\ref{B3}, \ref{complmomentum}, \ref{auxE}) with $6$
\end{enumerate}
all equivalent to the Dirac equation and where the last number indicates how many redundant equations each set has.

These $15$ groups, added to the $4$ groups seen before, amount to a total of $19$ groups that are equivalent to the Dirac equation. And they are the smallest sets to be equivalent to the Dirac equation, in the sense that in none of them we can remove an equation without losing the equivalence to the Dirac equation in spite of the fact that there might be some component of a covariant equation that is redundant. For instance, in the set (\ref{A1}, \ref{A3}, \ref{forwardmomentum}) the equations given by $E_{3}\!+\!P_{0}\!=\!0$, $F_{2}\!-\!P_{1}\!=\!0$ and $F_{1}\!+\!P_{2}\!=\!0$ are redundant but removing them would mean to remove the equation \eqref{forwardmomentum} and therefore also the equation $E_{0}\!+\!P_{3}\!=\!0$ which is not redundant so that the equivalence to the Dirac equation will be lost. While all above sets are the smallest in this sense, this is not a property of any possible set. For example, the set given by the divergences and curls of velocity and spin (\ref{A1}, \ref{B1}, \ref{A3}, \ref{B3}) is equivalent to the Dirac equation with $6$ redundant equations but the two equations \eqref{A1} and \eqref{B1} are fully redundant and they could be removed altogether leaving a set that is still equivalent to the Dirac equation. The largest would be to consider all equations obtaining a set with $24$ redundant equations. There is of course no need to do that, and hence having selected the $19$ smallest sets is a measure of the fact that the work has been done accurately enough to maintain a certain degree of simplicity.

The importance of selecting such sets can be seen from the fact that some of them might be remarkably easy to be interpreted. For instance, (\ref{A3}, \ref{B3}) consists of only curls of velocity and spin. Alternatively, (\ref{A1}, \ref{A3}, \ref{forwardmomentum}) has the curl and divergence of the velocity plus \eqref{forwardmomentum} providing the momentum \cite{Fabbri:2022kfr}. This easier visualization might be essential for the Broglie-Bohm approach to relativistic versions with spin of quantum mechanics \cite{Fabbri:2022kfr}.
\section{Conclusion}
In this paper, we have considered the general geometric construction of spinor fields re-formulated in the context of the polar variables. After extending the formulation to all differential structures with the introduction of the tensorial connection, we have re-written the Dirac differential field equations in hydrodynamic form. In doing this, we acquired the possibility to see that the Dirac equations are equivalent to the $4$ groups (\ref{A1}-\ref{A2}-\ref{A3}), (\ref{B1}-\ref{B2}-\ref{B3}), (\ref{forwardmomentum}-\ref{complmomentum}), (\ref{auxF}-\ref{auxE}), but also to the $8$ groups (\ref{A1}-\ref{A3}-\ref{forwardmomentum}), (\ref{A2}-\ref{A3}-\ref{complmomentum}), (\ref{A2}-\ref{A3}-\ref{auxF}), (\ref{A1}-\ref{A3}-\ref{auxE}), (\ref{B1}-\ref{B3}-\ref{forwardmomentum}), (\ref{B2}-\ref{B3}-\ref{complmomentum}), (\ref{B2}-\ref{B3}-\ref{auxF}), (\ref{B1}-\ref{B3}-\ref{auxE}) up to three redundant components of the field equations, to the $4$ groups (\ref{A3}-\ref{forwardmomentum}-\ref{auxF}), (\ref{A3}-\ref{complmomentum}-\ref{auxE}), (\ref{B3}-\ref{forwardmomentum}-\ref{auxF}), (\ref{B3}-\ref{complmomentum}-\ref{auxE}) up to six redundant components, to the $2$ groups (\ref{A1}-\ref{B1}-\ref{forwardmomentum}-\ref{auxE}), (\ref{A2}-\ref{B2}-\ref{complmomentum}-\ref{auxF}) up to two redundant components, and to the $1$ group (\ref{A3}-\ref{B3}) up to four redundant components, for a total of $19$ groups of field equations constituting the minimal sets, in the sense that no covariant equation can be removed by any of them without losing equivalence.

All these $19$ sets are the smallest to be equivalent to the Dirac equations, converting the quantum mechanics of the spinning relativistic particle into hydrodynamic form, the only form that is well-suited to render quantum mechanics visually interpretable, in the spirit that was first followed by de Broglie and Bohm. So what next?

There are two main things that would have to be done to bring this treatment to an acceptable level of completion, the first of which being the fact that writing everything in hydrodynamic form, treating the spinor field as one special type of fluid, does not in itself imply consistency. In fact, that specific type of fluid may still have problems, as pointed out in \cite{Florkowski:2018fap, Montenegro:2020paq, Weickgenannt:2022zxs}. More has to be done to make sure that, seen as a special fluid, the spinor field is well defined.

The second problem is also a problem of the de Broglie-Bohm interpretation, and that is we still do not know how exactly we can treat multi-particle systems in relativistic contexts, and more generally, we do not know how to make a second-quantized version of it. This problem is open since the 1960s and we are not trying to solve it here.

We hope that with such a variety of formulations, some problems may become of easier solution.

\end{document}